\documentclass[11pt,a4paper]{article}
\usepackage{jcappub}

\usepackage{amssymb}
\usepackage{amsthm}
\usepackage{amsmath}
\usepackage{amssymb}
\usepackage[utf8]{inputenc} 
\usepackage{url} 

\title{Quantum Big Bounce of the isotropic Universe via a relational time}
\author[a]{Eleonora Giovannetti,}
\author[a]{Fabio Maione,}
\author[b,a]{Giovanni Montani}
\affiliation[a]{Department of Physics, “Sapienza” University of Rome,\\P.le Aldo Moro, 5, Roma, Italy}
\affiliation[b]{ENEA, Fusion and Nuclear Safety Department, C.R. Frascati,\\
Via E. Fermi, 45, Frascati (RM), 00044, Italy}
\emailAdd{eleonora.giovannetti@uniroma1.it}
\abstract{We analyze the canonical quantum dynamics of the isotropic Universe in a metric approach by adopting a self-interacting scalar field as relational time. When the potential term is absent we are able to associate the the expanding and collapsing dynamics of the Universe with the positive and negative frequency modes that emerge in the Wheeler-DeWitt equation. On the other side, when the potential term is present a non-zero transition amplitude from positive to negative frequency states arises, as in the standard relativistic scattering theory below the particle creation threshold. In particular, we are able to compute the transition probability for an expanding Universe that emerges from a collapsing regime both in the standard quantization procedure and in the polymer formulation. The probability distribution results similar in the two cases and its maximum takes place when the mean values of the momentum essentially coincide in the in-going and out-going wave packets, as it would take place in a semiclassical Big Bounce dynamics.}
\begin{document}
\maketitle
%
%
%
%

\section{Introduction}	
The most relevant and long standing question of Relativistic Cosmology is surely the presence of an initial singularity in the dynamics of the isotropic Universe, as a general feature of the Einstein equations under cosmological hypotheses \cite{LL,PC}. Thought as a shortcoming of the underlying theory, the initial singularity has been the subject of a wide number of attempts devoted to its removal and originally focused on modifications of the Einstein-Hilbert action, able to alter the Friedmann dynamics of the Universe \cite{B1,B2,O}. Since the canonical quantization of gravity was initially formulated \cite{DeWitt67,PC}, the dominant proposal is the possibility that quantum effects at the Planck age could prevent the zero-volume limit characterizing the singularity. In this respect, understanding that in Loop Quantum Gravity (LQG) 	\cite{PC,CQG,LQG1,LQG2,LQG3} the volume operator has a discrete spectrum \cite{RS} and then showing the emergence of the Big Bounce semiclassical dynamics in the so-called Loop Quantum Cosmology (LQC) theory \cite{ashtekar2003,Ashtekar2006,AshtekarI,Ashtekar2008,Ashtekar2009,Ashtekar2011, Bojowald2002,Bojowald2004} was the most relevant success. Also Polymer Quantum Cosmology (PQC) is related to this scenario \cite{Pol,review}, due to its several morphological features in common with the quasi-classical limit of LQC. However, all these approaches are based on a semiclassical representation of the Universe dynamics, in which the Big Bounce is the consequence of a regularization of the standard Friedmann evolution of the Universe near the Planckian era. In this regard, in \cite{QB} it is underlined that a quasi-classical representation for the primordial phase is not always allowed since the Universe is in a fully quantum state. So, it is introduced the concept of a transition amplitude from a collapsing to an expanding Bianchi I Universe in the Wheeler-DeWitt (WDW) formulation, where the isotropic Misner variable plays the role of time. The associated transition probability results well-defined and with a Gaussian-like distribution. 
	
In this work we follow the same spirit of \cite{QB}, limiting our attention to the Friedmann-Lema\^{i}tre-Robertson-Walker (FLRW) Universe but generalizing the quantum formulation towards the implementation of a real matter field as relational time, here a self-interacting scalar field \cite{Rovelli}. Firstly, we construct a parallelism between the WDW equation and a Klein-Gordon (KG) one by identifying the expanding and the collapsing Universe as positive and negative frequency solutions respectively. Then, following the relativistic scattering procedure (below the particle creation threshold) \cite{BjD} we calculate the transition amplitude from negative to positive frequency states, which turns out in a \textquotedblleft Quantum Big Bounce" picture. The probability associated to this transition is maximum when the momentum mean value of the in-going Universe wave packet is nearly that one of the out-going one. Finally, we consider the same model in Polymer Quantum Mechanics (PQM), which is not able to remove the singularity in the associate WDW equation when written in the Misner variables (see \cite{Crino,GM} when the singular behavior of the Bianchi I, and hence of the Bianchi IX model, is outlined). However, the polymer paradigm singles out a spreading feature which prevents a semiclassical description, making the concept of a Quantum Big Bounce as more robust. In consideration of the study developed in \cite{P2}, the surprising result is that the profile of the probability density is essentially unchanged in the polymer formulation with respect to the standard quantum scheme. Basically, the present analysis suggests that a Quantum Big Bounce is allowed at a probabilistic level in the WDW formalism for the isotropic Universe, thanks to a proper interpretation of the WDW solutions as negative and positive frequency states defined in terms of a matter relational time. We remark that the possibility to have a transition from a collapsing to an expanding Universe is due to the breaking of the frequency separation because of the presence of a potential energy density in the dynamics of the matter clock and that the position of the peak in the probability density reflects the same symmetrical reconnection of the singular branches typical of a semiclassical bouncing picture \cite{PC}. 
\section{\label{sec:level1} Quantum FLRW dynamics}
The isotropic Universe is described by the FLRW model and its classical dynamics is affected by past or future singularities, i.e. the Big Bang and the Big Crunch respectively. The Hamiltonian of the system in the presence of a free scalar field $\phi$ is
\begin{equation}
\label{H}
H =N\mathcal{H}=N\left[-\frac{2}{3} \frac{\pi G}{c^3V} p_{\alpha}^2 +\frac{c}{2V} p_\phi^2 \right]=0\,,
\end{equation}
where $N$ is the lapse function, $\alpha=\ln{a}$ is the isotropic Misner variable, $G$ is the Einstein constant, $c$ is the speed of light in vacuum and $V$ is the fiducial volume.
In order to preserve the covariant structure of the theory, we use the so-called Dirac procedure and quantize the theory before having explicitly solved the constraints, now promoted to quantum operators. So, implementing the super-Hamiltonian constraint as a quantum operator $\hat{\mathcal{H}}$ that annihilates the Universe wave function $\Psi(\alpha,\phi)$ we obtain the WDW equation
\begin{equation} \label{WDW1}
\hat{\mathcal{H}}\Psi(\alpha,\phi)=\frac{2}{3} \frac{\pi G\hbar^2}{c^3 V } \frac{\partial^2\Psi(\alpha,\phi)}{\partial \alpha^2}  -  \frac{c\hbar^2}{2V} \frac{\partial^2\Psi(\alpha,\phi)}{\partial \phi^2}= 0\,,
\end{equation}
where we have used the coordinate representation and the normal ordering convention ($\hbar=h/2\pi$ is the Planck constant). We note that equation \eqref{WDW1} is formally analogous to a two-dimensional massless KG equation for a relativistic particle. Thanks to this analogy, we can identify $\phi$ as a relational time variable and $\alpha$ as the spatial degree of freedom. Hence, according to the superposition principle, the most general solution is a linear combination of plane waves, i.e. a wave packet
\begin{align}
\label{solgen}
&\psi (\alpha, \phi) = \int_{-\infty}^{+\infty} [A_+(k) \varphi_+(\alpha,\phi)+A_-(k) \varphi_-(\alpha,\phi)]\,dk\,,\\
&\varphi_{\pm}(\alpha,\phi)=e^{i(k\alpha\mp \omega_k\phi)}\,,
\end{align}
where $k$ is the wave number, $\omega_k = v |k| $ is the dispersion relation (i.e. the frequency), $A_{\pm}(k)$ are arbitrary functions and $v=\sqrt{4\pi G\hbar^2/3c^4}$.

Actually, when dealing with highly-localized wave packets it is possible to relate the classical values of the momenta (i.e. $p_\alpha$ and $p_\phi$) and the corresponding quantum eigenvalues (i.e. $k$ and $\omega_k$ respectively), in agreement with the Ehrenfest theorem. In this respect, we note that by deriving the Hamilton equations
\begin{subequations}
\begin{align}
\label{phi}
&\dot{\phi} = \frac{\partial H}{\partial p_\phi}\frac{Ne^{-3\alpha}c p_\phi}{V}\,,\hspace{0.5cm}\dot{p}_\phi = -\frac{\partial H}{\partial \phi} =0\,,\\
&\dot{\alpha} = -\frac{Ne^{-3\alpha}4\pi Gp_\alpha}{3c^3 V }\,,\hspace{0.3cm}
\dot{p}_\alpha = 3H=0
\end{align}	
\end{subequations}	
from \eqref{H} and by combining the equations for $\dot{\alpha}$ and $\dot{\phi}$ we obtain the relation
\begin{equation}\label{dalfadphi}
\frac{d\alpha}{d\phi} = -\frac{4\pi G}{3c^4}\frac{p_\alpha}{p_\phi}
\end{equation}
that has a covariant character, i.e. it is valid for any temporal gauge. From \eqref{dalfadphi} we get that the distinction between the classical branches depends on the relative sign between the constant of motion $ p_\alpha $ and $ p_\phi $: the Universe collapses if the sign is concordant, otherwise it expands. In particular, once fixed the sign of $p_{\alpha}$ the expanding or collapsing feature depends on the sign of $p_{\phi}$. Therefore, if we consider sufficiently localized (i.e. semiclassical) wave packets far from the fully quantum region of the initial singularity, we can select only the states with $k>0$ from the full Hilbert space and infer that the positive frequency solutions of the type
\begin{equation}
\label{psi+}
\psi_{+}(\alpha,\phi) =\int_{0}^{+\infty} A_{+}(k) e^{i(k\alpha-\omega_k\phi)}\,dk
\end{equation}
describe an expanding Universe, whereas the negative frequency ones
\begin{equation}
\label{psi-}
\psi_{-}(\alpha,\phi) =\int_{0}^{+\infty} A_{-}(k) e^{i(k\alpha+\omega_k\phi)}\,dk
\end{equation}
a collapsing one. Thus, we can associate positive energy states (i.e. \textquotedblleft particles states") with the expanding branch and negative energy ones (i.e. \textquotedblleft antiparticles states") with the collapsing phase. We remark that considering negative values of $p_\alpha$ leads to the opposite identification. Actually, retaining both signs of $p_{\alpha}$ would provide a redundant information. So, in \eqref{psi+} and \eqref{psi-} we choose the portion of the spectrum of the operator $\hat{p}_{\alpha}$ associated to $k>0$ and we impose $A_{\pm}(k)=e^{-\frac{(k-\overline{k})^2}{2\sigma^2}}/\sqrt{2\pi \sigma^2}$, in which $ \bar{k} $ and $ \sigma $ are fixed by the initial condition on the wave function at a given $ \phi = \phi_0 $.

We remark that the frequency separation can be performed only in the absence of a self-interacting potential of the scalar field (here a time-dependent term) and that the choice of the semi-axis of $k$ on which performing the integration is conventional. We also stress that the pure classical dynamics in function of time $\phi$ would require a fixed sign for $p_\phi$ and then the two branches would depend on the Cauchy condition on the sign of $p_\alpha$. Actually, in this picture we use a \textquotedblleft time after quantization" approach thanks to the analogy with the KG equation. Hence, in this fully quantum scenario we have a pure relational dynamics in which states can propagate forward and backward in time as in the relativistic quantum particles interactions. Furthermore, approaching the classical limit the emergent state is that of an expanding Universe with fixed (here negative) $p_\phi$ and therefore the consistency of the classical dynamics is preserved. In addition, from \eqref{phi} we can justify \emph{a posteriori} the viability of $\phi$ as a relational time variable. In fact, $p_\phi $ is a constant of motion and so $\phi$ is a monotonic function of the synchronous time. In this respect, the advantage of using a matter relational time is that the monotonicity requirement would be fulfilled also in a bouncing picture.
 
\section{\label{sec:level11} Quantum FLRW dynamics with a self-interacting scalar field}
In this section we introduce a potential term $U(\phi)$ at a quantum level. We note that the WDW approach is not able to regularize the dynamics of the present model. However, the proposed approach has the purpose to demonstrate the presence of the Big Bounce in the WDW formulation by treating it as a quantum scattering. Hence, we consider the quantum potential $U(\phi)= \frac{\lambda}{2}e^{-n\phi}$, where $\lambda > 0$ and $n \in \mathbb{R}$ (for cosmological implementations of such a potential, see \cite{staro,quint1,quint2}). So, the WDW equation turns out to be
\begin{equation}
\label{WDWU}
\left[\frac{\partial^2}{\partial\alpha^2}  - \frac{1}{v^2} \frac{\partial^2}{\partial\phi^2} + Ce^{6\alpha-n\phi}\right]\psi(\alpha,\phi) = 0\,,
\end{equation}
where $C= 3\lambda(V c)^2/4\hbar^2\pi G$ ($v$ can be taken out from the equation by redefining $\bar{\phi} = v\phi$ and $\bar{n}=n/v$ and then renaming the new variables as the old ones). As before, \eqref{WDWU} can be interpreted as a KG equation where the potential term depends on both the spatial and time variable. To solve the equation, we use the transformation
\begin{equation}	
\label{newv}
\begin{cases}
\alpha -\sqrt{2}\phi = -\eta\\
\sqrt{2}\alpha -\phi = \xi
\end{cases}
\end{equation}
that has the properties of a proper Lorentz transformation in the Minisuperspace. Thus, the kinetic term remains diagonal and the WDW equation in the new variables becomes
\begin{equation} \label{newWDW}
\left[ \frac{\partial^2}{\partial\xi^2}  -  \frac{\partial^2}{\partial\eta^2} + Ce^{-6\eta}\right]\Psi(\xi,\eta) = 0\,.
\end{equation}
For a detailed discussion on the issue of selecting the unitarily-equivalent physical solutions at a quantum level in constrained systems, see \cite{WDWrules}.

It is worth noting that the change of variables \eqref{newv} mixes the spatial and time coordinates, making their immediate interpretation difficult. Recalling the hypothesis that the introduced potential is time-dependent, $ \eta $ can be interpreted as the time variable and $ \xi $ as the spatial one. We note that this potential is relevant only near the singularity. By analyzing the Hamilton equations for the new variables we obtain $\frac{d\xi}{d\eta} = -\frac{p_\xi}{p_\eta}$, so the classical relation between the new spatial and time coordinates is the same as \eqref{dalfadphi}. This is due to the fact that the proper Lorentz transformations do not invert the arrow of time. Therefore, in the new variables it remains valid that the positive frequency solutions can be associated to an expanding Universe, whereas the negative frequency ones to a collapsing Universe. 

Now, we propose the solution
\begin{equation}
\Psi^k(\xi,\eta) = \Phi^k(\eta) \psi^k(\xi)
\end{equation}
to \eqref{newWDW}, where $ \psi_k (\xi) = e^{ik \xi} $ and
\begin{align}
 \nonumber 
 \Phi^k(\eta)&= \Phi^k_+(\eta)+\Phi^k_-(\eta)=\\ \nonumber &=6^{\frac{-ik}{3}}\mathcal{B}_I\big(-ik/3,e^{-3\eta}\sqrt{C}/3\big)\Gamma\big(1-ik/3\big) +\\ 
 &+6^{\frac{ik}{3}}\mathcal{B}_I\big(ik/3,e^{-3\eta}\sqrt{C}/3\big)\Gamma\big(1+ik/3\big)\,,
 \label{soluzione I}
\end{align}	
in which $\mathcal{B}_I$ are the modified Bessel functions of the first kind and $\Gamma$ is the Euler gamma function. As before, we write the general solution $ \Psi(\xi,\eta) $ by means of Gaussian packets
\begin{equation}
\label{Psi}
\Psi(\xi,\eta) =\frac{1}{\sqrt{2\pi \sigma^2}} \int_{-\infty}^{+\infty} e^{-\frac{(k-\bar{k})^2}{2\sigma^2}} \Phi^k(\eta)e^{ik\xi}\,dk\,.
\end{equation}
Differently from the previous case, here the potential term has mixed the branches, making it impossible to distinguish them in terms of frequencies. Moreover, it is no longer possible to find an explicit expression for the eigenvalue of the operator $ \hat{p}_\eta $ in terms of the eigenvalue of $ \hat{p}_\xi $, since the separation constant $ k $ appears as a complex index of the Bessel functions. 
We finally note that the Bessel functions tend to the free solutions in the limit $ \eta \rightarrow +\infty $ , i.e.
\begin{equation}
\lim_{\eta \rightarrow +\infty}\Phi^k_+(\eta) = e^{ik\eta} \quad \text{and} \quad\lim_{\eta \rightarrow +\infty} \Phi^k_-(\eta) = e^{-ik\eta}\,.
\end{equation}
In other words, when the time-dependent potential becomes negligible $\Psi^k(\xi,\eta)$ is that of the free case and therefore we recover the picture described in the previous section. In particular, in \eqref{soluzione I} the two different asymptotic behaviours (i.e. expanding and collapsing) are equally weighed. 
	
\section{\label{QB} Quantum Big Bounce}
\begin{figure}[h]
\centering
\includegraphics[scale=0.6]{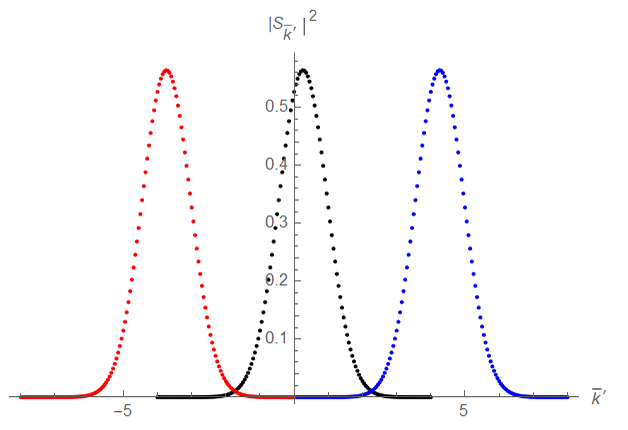}
\caption{Plot of $|S(\bar{k},\bar{k'})|^2$ as function of $ \bar{k}'$. Starting from the left, the transition probability has been calculated for $ \bar{k}= -4,0,4$ respectively. In all the three cases $ \sigma = \sigma' = 1 $, $ C = 1 $ and an integration step equal to 0.05 has been used.}
\label{fig:ampiezzaI}
\end{figure}

Here we propose a probabilistic approach to the Big Bounce within the WDW theory. At the basis of the formalism used below there is the analogy with the relativistic quantum scattering theory. In the same spirit, we assume that the Bounce is a quantum interaction between single particle states, i.e. they can be described by a wave function. In this framework, we associate a probability to the transition of the Universe from a collapsing state to an expanding one, in analogy with the fundamental interaction processes between particles below the particle creation threshold \cite{BjD}. In the presence of a potential the general solution of the following modified KG equation
\begin{equation}
(\Box_x + m^2 + U(x)) \Psi(x,t) = 0
\end{equation}
can be found in terms of the propagator for the free scalar field and calculated with arbitrary precision as
\begin{equation}
\Psi(x,t) = \Phi(x,t) + \int \Delta_F(x-y) U(y) \Psi(y,t)\,d^4 y \,,
\end{equation}
where $ \Phi(x,t) $ is the solution of the homogeneous problem and $ \Delta_F(x-y) $ is the propagator, which has the property of making the positive frequency solutions evolve forwards in time and the negative frequency ones backwards. In this case, the probability associated to the transition amplitude can be written as
\begin{equation}\label{modulo quadro}
|S_{\bar{k}',\bar{k}}|^2 =\bigg|-i \iint_{-\infty}^{+\infty} \psi^*_+(\xi,\eta)  U(\eta)\Psi(\xi,\eta)\,d\eta\,d\xi  \bigg|^2\,,
\end{equation}
where $ \Psi(\xi,\eta) $ coincides with the solution resulting from the interaction of a collapsing Universe with the potential $U(\eta)=C e^{-6\eta}$ (see \eqref{Psi}) and $\psi^*_+(\xi,\eta)$ is an emerging free wave associated only to positive frequencies
\begin{equation}
\psi^*_+(\xi,\eta)=  \frac{1}{\sqrt{2\pi \sigma'^2}}\int_0^{+\infty}\frac{e^{-\frac{(k'-\overline{k}')^2}{2\sigma'^2}}}{\sqrt{2w_{k'}}}e^{i(k'\xi\hspace{0.5mm} - \hspace{0.5mm} w_k'\eta)}\,dk'\,,
\end{equation}
i.e. the branch of the expanding Universe. We remark that the probability amplitude in \eqref{modulo quadro} refers to a transition from a negative (collapsing) to a positive (expanding) frequency state by definition. Actually, the scattering amplitude between two \textquotedblleft particles" or \textquotedblleft antiparticles" would have contained a Dirac delta term added to the integral in \eqref{modulo quadro} (see \cite{BjD}). After both analytical and numerical integrations, the transition probability $|S(\bar{k},\bar{k}')|^2$ results to be dependent only on the values $(\bar{k}, \bar{k}') $ associated to the in-going (collapsing) and out-going (expanding) wave packet respectively. So, once fixed the value of $ \bar{k} $, the result is a function of $ \bar{k}' $, as shown in Fig. \ref{fig:ampiezzaI}.
We notice that the transition probability between a collapsing and an expanding Universe is well-defined. In particular, its peak position indicates that the wave packet with $ \bar{k}'\approx \bar{k} $ maximizes the probability of the Bounce. It is worth noting that the phase shift observed in Fig. \ref{fig:ampiezzaI} between $ \bar{k}' $ and $ \bar{k} $ is due to a mathematical feature of the Bessel functions and it has no physical meaning. Finally, we stress that the KG probability density becomes positive-defined only when the frequency separation is performed (i.e. when the potential term is negligible). Surprisingly, this feature holds also when we overlap frequency modes of the same sign by using Gaussian weights (for a detailed discussion of this question, see \cite{QB}). However, when a time-dependent potential is turned on the frequency separation is forbidden and the KG norm is not globally conserved anymore, so its interpretation as a probability loses its meaning at all. Actually, the approach we adopt here goes beyond the troublesome probabilistic interpretation of the KG wave function by resorting to the relativistic quantum formalism of scattering processes described above. Here, the probability associated to the Quantum Big Bounce is simply the square modulus of the transition amplitude constructed by projecting the in-going state onto the out-going one, according to the KG scalar product. Hence, the positive nature of the Quantum Big Bounce probability density as defined in \eqref{modulo quadro} is automatically guaranteed. In this respect, the present construction is a natural implementation of the relativistic quantum theory procedures into the Minisuperspace of the considered cosmological model.
\begin{figure*}
\begin{minipage}[h]{3.4cm}
\centering
\includegraphics[width=1.7\linewidth]{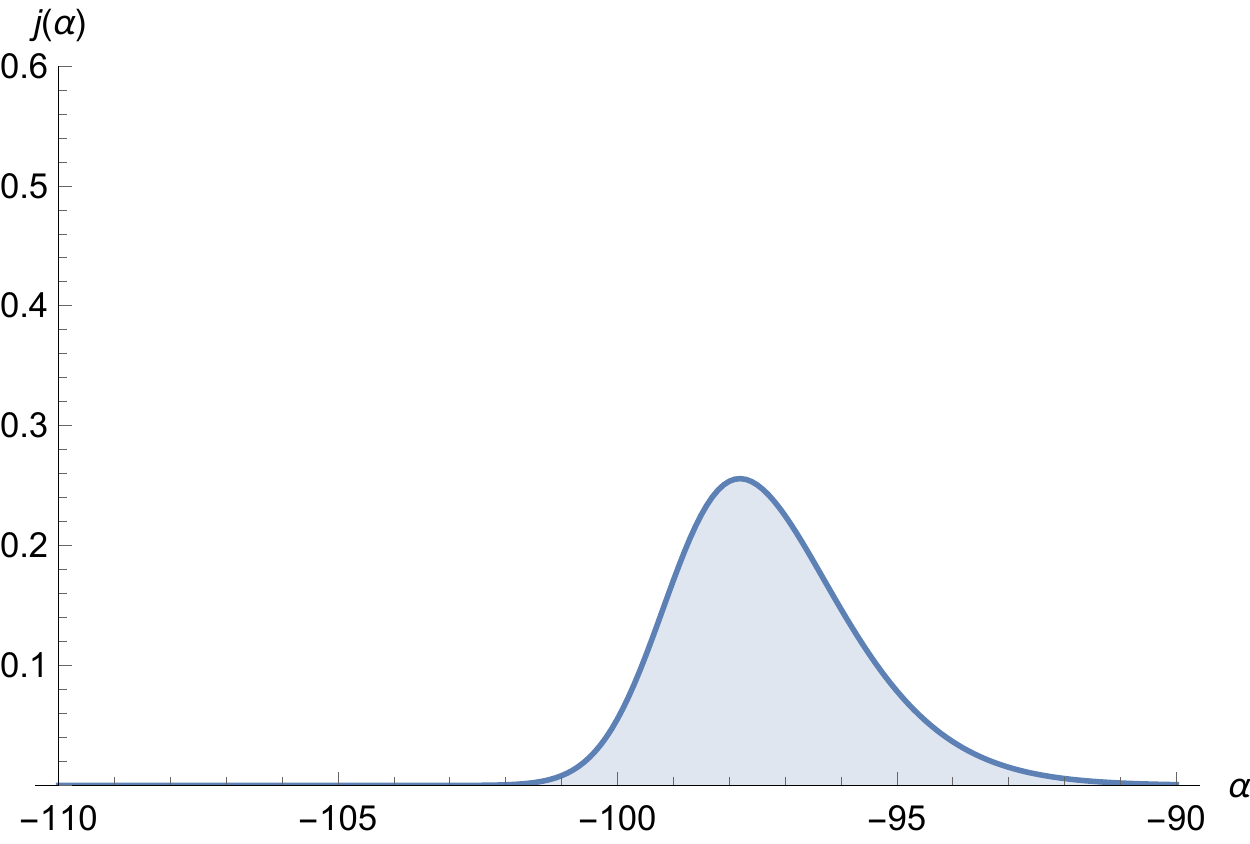}
\end{minipage}\hspace{2.5cm}
\begin{minipage}[h]{3.4cm}
\centering
\includegraphics[width=1.7\linewidth]{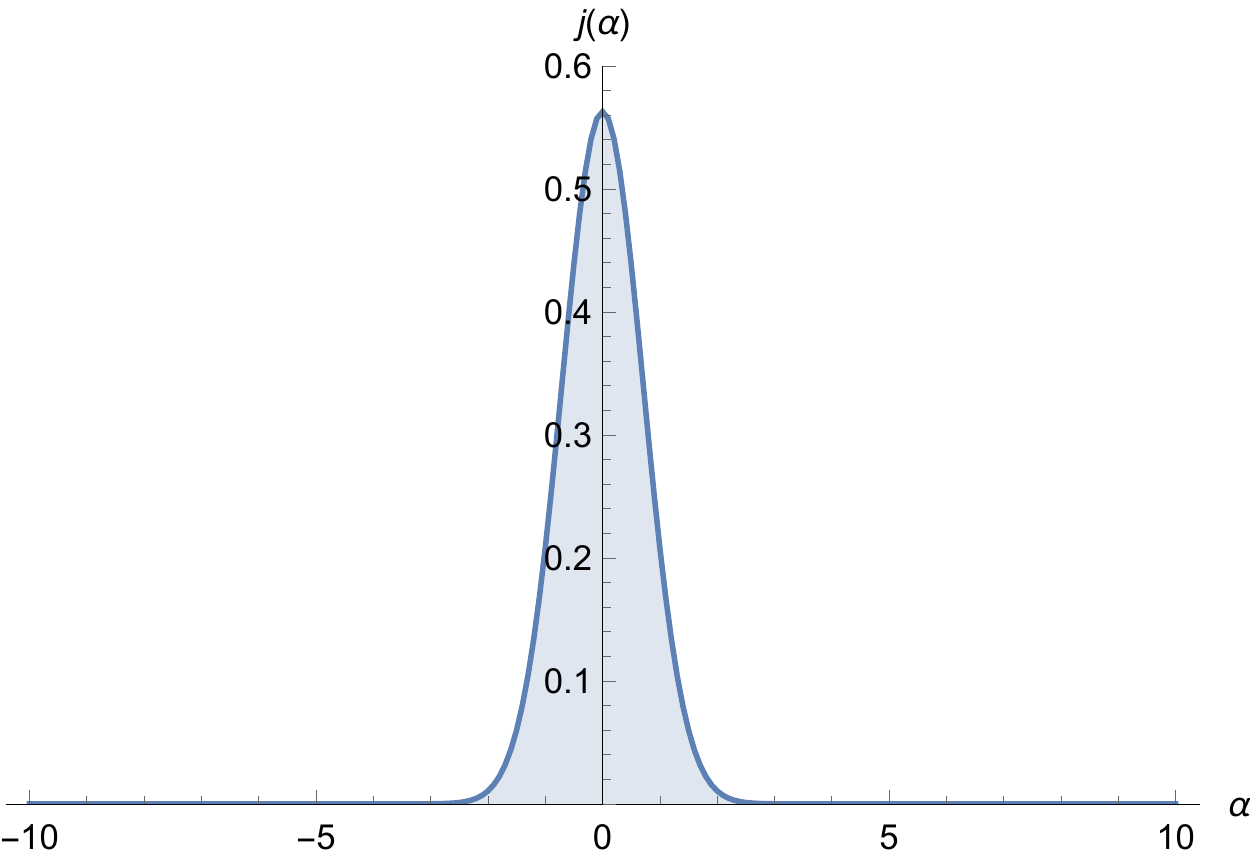}
\end{minipage}\hspace{2.5cm}
\begin{minipage}[h]{3.4cm}
\centering
\includegraphics[width=1.7\linewidth]{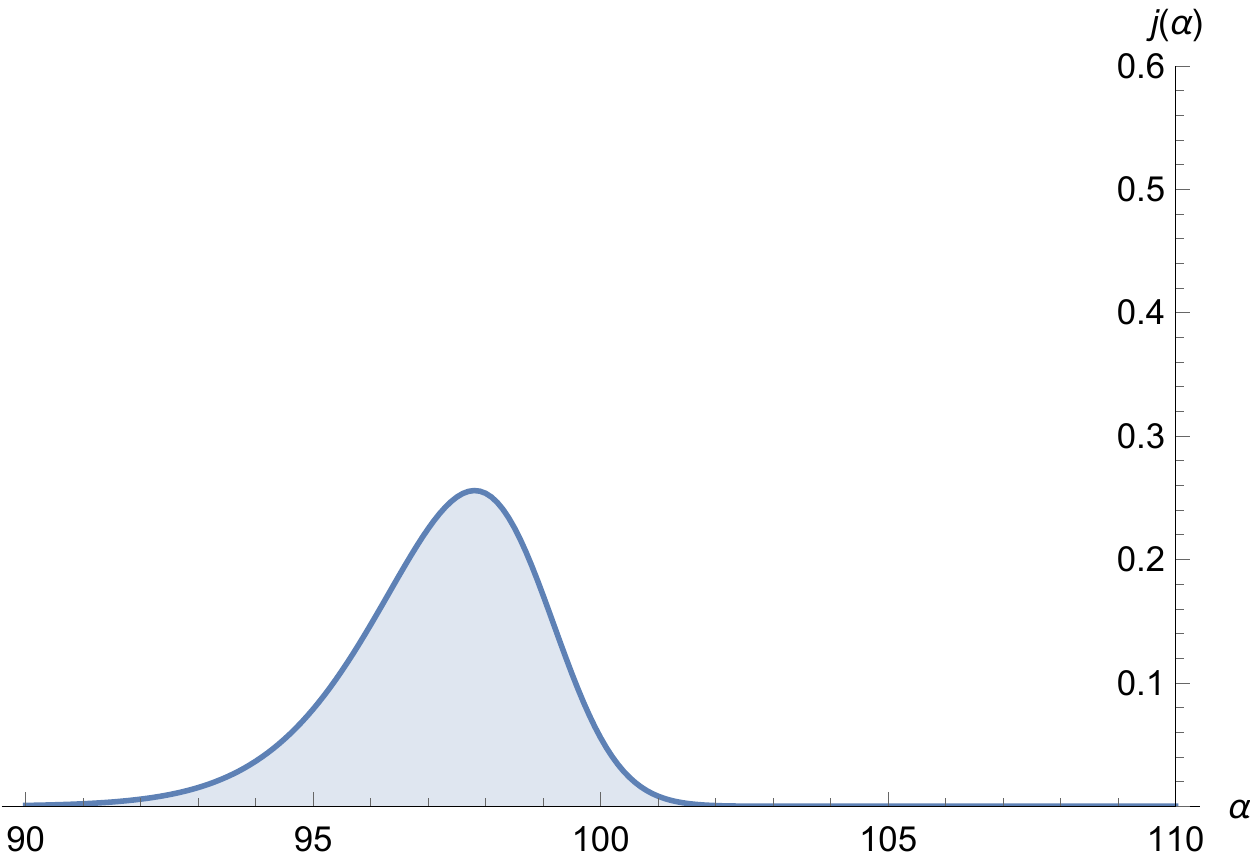}
\end{minipage}\hspace{2.5cm}
\caption{Probability density $j(\alpha)=i(\psi_+^*(\alpha,\bar{\phi})\partial_\phi\psi_+(\alpha,\bar{\phi})-\psi_+(\alpha,\bar{\phi})\partial_\phi\psi_+^*(\alpha,\bar{\phi})) $ for the values of time $ \bar{\phi} =-100,0,100$ respectively. As it can be seen, the variance of the probability density becomes as larger as time goes, showing the spreading of the wave packet during the evolution. Both the integration step and $ \mu_\alpha$ have been set equal to 0.1.}
\label{spread}
\end{figure*}

\section{\label{sec:leve222} Quantum Big Bounce in the polymer paradigm}
In the last part of the work we study the system by implementing the PQM formalism \cite{Pol}. The ultimate goal is to make a comparison with the transition amplitude obtained in the previous formalism and analyze if there are significant variations in the two cases. First of all, in accordance with the theoretical structure of PQM it is necessary to select which Minisuperspace variables are discrete and therefore which ones should be regularized. When PQM is applied in a cosmological context, functions of the scale factor are usually chosen as discrete with the aim of trying to solve the singularity. Accordingly, in this model $ \alpha $ is the discrete variable and so the operator $ \hat{p}_\alpha $ is formally replaced by
\begin{equation}
\label{pol}
\hat{p}_\alpha \rightarrow \frac{\hbar}{\mu_\alpha}\sin\left(\frac{\mu_\alpha p_\alpha }{\hbar} \right)
\end{equation}
in the momentum representation, where the polymer parameter $ \mu_\alpha $ verifies the condition $\mu_\alpha p_\alpha \ll \hbar$. From the semiclassical study of the Hamilton equations in the absence of the potential term we get that $p_\alpha$ is still a constant of motion and that
\begin{equation} \label{eq moto polymer}
\dot{\alpha} = -\frac{4\pi G Ne^{-3\alpha}\hbar}{3c^3V\mu_\alpha}\sin\left(\frac{2\mu_\alpha p_\alpha}{\hbar} \right)\,.
\end{equation}
Hence, it is clear that the same previous considerations regarding the possibility of discerning the expanding and collapsing branches of the Universe remain valid, since $\sin(2\mu_\alpha p_\alpha/\hslash) $ is odd in its argument.
As usual, the polymer version of the Hamiltonian \eqref{H} is obtained by using \eqref{pol} as a semiclassical approximation. Then, promoting it to a quantum operator we obtain the polymer-modified WDW equation, that corresponds to \eqref{WDW1} in which \eqref{pol} is rigorously implemented as a regularized version of the operator $\hat{p}_\alpha$ in the momentum representation. The expression of the Universe wave packet as a superposition of functions weighed with Gaussian coefficients is analogous to the previous one (see \eqref{Psi})
\begin{equation}
\psi_\pm(\alpha,\phi) = \frac{1}{\sqrt{2\pi\sigma^2}}\int_0^{+\infty} e^{-\frac{(k-\bar{k})^2}{2\sigma^2}} e^{i(\bar{p}_\alpha \alpha \mp k\phi )}\,dk\,,
\end{equation}
where $\bar{p}_\alpha =\arcsin(\mu_\alpha k)/\mu_\alpha$. The main difference is represented by the dispersion relation (i.e. the relation between the quantum eigenvalues of the momenta) which stops being linear. It can be shown that this fact is responsible for the spreading of the wave packet (see Fig. \ref{spread}). Therefore, the comparison between the semiclassical trajectory and the evolution of the quantum wave packet loses its meaning since the variance sooner or later becomes of the same order as the expectation value, making it necessary to resort to a fully quantum treatment. This behavior of the FLRW Universe wave packet in the polymer representation further consolidates the study of the Big Bounce as a quantum scattering. We remark that PQM is not able to solve the singularity of the FLRW model in the Misner variable (see \cite{Crino,GM} where this argument is used to demonstrate the presence of the initial singularity in the Bianchi IX model) and so no semiclassical bouncing dynamics is present in this model.  
	
\begin{figure}[h]
\centering
\includegraphics[scale=0.6]{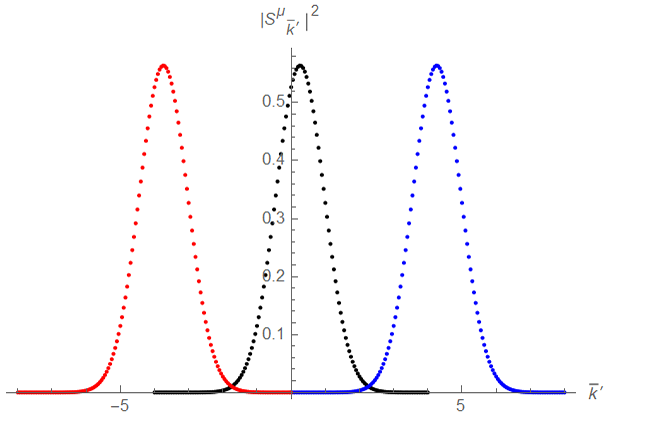}
\caption{Plot of $|S^{\mu_\xi}(\bar{k'},\bar{k})|^2$ as function of $ \bar{k}'$ in the polymer paradigm. Starting from the left, the polymer transition probability has been calculated for $ \bar{k}=-4,0,4$ respectively. In all the three cases $ \sigma = \sigma' = 1 $, $  C = 1, \mu_\xi = 0.1 $ and an integration step equal to 0.05 has been used.}
\label{fig:ampiezzaPolymer}
\end{figure}

Since PQM does not commute with other coordinate transformations, we introduce it after the Lorentz transformation \eqref{newv}. Hence, we start by considering the polymer version of the WDW equation in the $ (\xi,\eta) $ variables, i.e.
\begin{equation}
\left[ \frac{1}{\mu_\xi^2}\sin^2\left( \mu_\xi p_\xi\right) -\frac{\partial^2}{\partial\eta^2 } +Ce^{-6\eta} \right] \Psi(p_\xi,\eta) = 0\,,
\end{equation}
where the substitution \eqref{pol} has been used.
Then, the polymer transition probability can be computed as in \eqref{}, in which
\begin{equation}
\psi^*_+(\xi,\eta)=  \frac{1}{\sqrt{2\pi \sigma'^2}}\int_0^{+\infty}\frac{e^{-\frac{(k'-\overline{k}')^2}{2\sigma'^2}}}{\sqrt{2w_{k'}}}e^{i(\bar{p}_\xi'\xi\hspace{0.5mm} - \hspace{0.5mm} w_k'\eta)}\,dk'\,,
\end{equation}
$U(\eta)=Ce^{-6\eta}$ and
\begin{equation}
\Psi(\xi,\eta) =\frac{1}{\sqrt{2\pi \sigma^2}} \int_{-\infty}^{+\infty} e^{-\frac{(k-\bar{k})^2}{2\sigma^2}} \Phi^k(\eta)e^{i\bar{p}_\xi\xi}\,dk\,.
\end{equation}
After analitically integrating over the variables $\xi$ and $k'$, we get
\begin{align}
\label{Spol}
&|S^{\mu_\xi}(\bar{k'},\bar{k})|^2=\\\nonumber&= \bigg|-i\int_{0}^{+\infty}\frac{Ce^{-\frac{(k-\overline{k})^2}{2\sigma^2}}e^{-\frac{(k-\overline{k}')^2}{2\sigma'^2}}}{{\sqrt{2w_{k}\sigma^2\sigma'^2}}}\sqrt{1-\mu_\xi^2k^2} I(k)\,dk \bigg|^2,
\end{align}
where 
\begin{equation}
I(k) = \int_{-\infty}^{+\infty} e^{\eta (ik -6)} \Phi^k(\eta)\,d\eta\,.
\end{equation}	
Then, the polymer transition probability $|S^{\mu_\xi}(\bar{k'},\bar{k})|^2$ has been computed by using both analytical and numerical integrations. We notice that the polymer modification induced on \eqref{Spol} consists in the presence of the global factor $\sqrt{1-\mu_\xi^2k^2}$ and that in the limit $\mu_\xi\rightarrow 0$ the standard scattering amplitude is recovered. Therefore,  as we can see from Fig. \ref{fig:ampiezzaPolymer} the introduction of PQM does not significantly change neither the shape of the transition probability $|S^{\mu_\xi}(\bar{k},\bar{k'})|^2$ nor the position of its peak (for a comparison with the previous case, see Fig. \ref{fig:ampiezzaI}). In particular, as in the standard case analyzed in Sec. \ref{QB} the maximum of the probability density occurs in correspondence of $\bar{k}' \approx \bar{k}$, where $\bar{k'}$ is associated to the expanding Universe wave packet emerging after the interaction and $ \bar{k} $ to the collapsing one. The absence of relevant effects is due to the fact that in \eqref{Spol} the exponential factors of the Gaussian coefficients dominate over the polymer global factor $\sqrt{1-\mu_\xi^2 k^2}$ in the integral.

\section{Discussion of the results}
Here we develop some basic considerations that are useful to provide a physical interpretation of the analysis above. We start by observing that the collapsing and expanding branches are separated on a classical level and no semiclassical dynamics is present, both in the Einsteinian and in the polymer picture addressed here. The situation is significantly different on a quantum level, where the expanding and collapsing branches co-exist as positive and negative frequency solutions. The crucial point is that the two frequency solutions, i.e. the collapsing and expanding branches, can no longer be separated when a self-interaction potential term (to be considered as relevant in the Planckian regime only) is introduced. As a result, the quantum nature of the singularity acquires a very different morphology from the standard expectancy according to which the trajectories are localized near the two singular branches. Actually, in the polymer picture a quasi-classical localization of the dynamics is possible for a finite time interval only and the solution describing the two branches should be regarded as a non-localized state near the singularity, even in the absence of a quantum potential term. Thus, here we state that it is possible to obtain an emerging classical expanding Universe for $\eta \rightarrow +\infty$ from the Planckian quantum state in which the frequency solutions are not separable. For the reasons explained above, a quantum transition between a mixed Planckian state and the expanding (but, in principle, also collapsing) Universe emerges as a new scenario that can be interpreted as a Quantum Big Bounce. We note that the precise morphology of a quantum scattering process would require the quantum potential to be a time transient, i.e. relevant only in a finite interval of $\phi$. Nevertheless, we stress that dealing with the exact solution of the interaction regime makes it possible to use the relativistic quantum treatment, even if such a potential term is not perturbative in this region. We conclude by observing that the present study must be regarded as an improvement of the original proposal in \cite{QB}, in which the Universe volume is adopted as time. Here, the choice of a relational time would intrinsically fulfil the requirement of a monotonic behavior even when considering a semiclassical/quantum bouncing dynamics.	
	
\section{Conclusions}
In this work we analyze the quantum dynamics of the isotropic Universe in the presence of a self-interacting scalar field which plays the role of a relational time \cite{Rovelli}. When the scalar field is free of its potential, the WDW equation has the morphology of a two-dimensional massless KG equation and we are able to identify the positive and negative frequency modes as the expanding and collapsing phases of the Universe respectively. Such an identification is found by comparing the classical dynamics with the behavior of localized quantum wave packets. Then, when the potential term of the matter clock is introduced, the frequency separation is broken and so we apply the standard techniques of relativistic quantum scattering below the particle creation threshold \cite{BjD}, in order to get a transition probability from a collapsing to an expanding Universe. It is worth noting that we use an orthochronous Lorentz-like transformation in the Minisuperspace when performing such a calculation, in order to make it possible to analytically solve the WDW equation. In both cases, i.e. the standard WDW equation and the polymer modified scheme, the transition probability results similar in morphology. Furthermore, in the latter scenario the Universe wave packet unavoidably spreads towards the (still existing) initial singularity and the Big Bounce description in terms of a probabilistic phenomenon becomes mandatory. Actually, in both cases the idea of dealing with a Quantum Big Bounce is made precise by observing that the maximum value of the transition probability is taken when the expectation value of the in-going wave packet momentum is close to the corresponding one in the out-going state, as it takes place before and after a semiclassical Big Bounce. Since the existence of a self-interacting scalar field in the early Universe appears natural in many fundamental approaches \cite{PC}, the possibility of dealing with a Quantum Big Bounce (in the sense defined here) can be considered as a rather general feature of the canonical quantum dynamics in the metric formalism, even in more general cosmological models like the Bianchi Universes or the generic inhomogeneous cosmological solution \cite{Benini_2004,Benini_2006}.

\bibliographystyle{elsarticle-num} 
\bibliography{bibi}

\end{document}